\documentstyle[11pt,aaspp4]{article}
\begin{document}

\title{CCD PHOTOMETRY OF FAINT VARIABLE STARS IN THE GLOBULAR 
CLUSTER NGC~6752
\footnote{
Based on observations collected at the Las Campanas Observatory of the
Carnegie Institution of Washington.
}}

\author{Ian B. Thompson}
\affil{Carnegie Observatories, 813 Santa Barbara St., Pasadena, CA 91101\\ 
e-mail: ian@ociw.edu}

\author{Janusz Kaluzny and Wojtek Pych}
\affil{Warsaw University Observatory, Al. Ujazdowskie ~4, 
00-478 Warsaw, Poland\\
e-mail: jka@vela.astrouw.edu.pl, pych@sirius.astrouw.edu.pl }

\author{Wojtek Krzeminski}
\affil{Las Campanas Observatory, Casilla 601, La Serena, Chile\\
email:wojtek@lco.cl}

\begin{abstract}
We present the results of a photometric survey for variable stars in
the field of the nearby globular cluster NGC~6752. The cluster was
monitored in 1996 and 1997 for a total of 54 hours with 3 different CCD
cameras mounted on the 1.0-m Swope telescope. Eleven new variables were
identified:  3 SX~Phe stars, 7 contact binaries and 1  candidate
detached eclipsing binary. All 3 SX Phe variables are likely members of
the cluster while only 1 out of the 7 contact binaries is a potential
cluster member.  As a by-product of our survey we obtained  $UBV$
photometry for a large sample of stars in the cluster field. Two stars
with $U-B\approx -1.0$ and $V=19.3$ and $V=20.6$ were identified. They
lie along the extended horizontal branch of the cluster, and are likely
to be faint sdB stars from NGC~6752.

\end{abstract}

\section{Introduction}

NGC~6752 is a medium-rich globular cluster whose proximity, low
reddening and relatively high galactic latitude ($r \approx 3.8$~kpc,
$E(B-V)=0.04$, $b=-25.6$; Harris 1996) make it an excellent object for
detailed studies. The cluster was selected as one of the targets of an
ongoing survey for eclipsing binaries in globular clusters (Kaluzny,
Thompson \& Krzeminski 1997).  The ultimate  goal of the project is to
use observations of detached eclipsing binaries to determine the ages
and distances of globular clusters (Paczy\'nski 1997), and to study the
binary star fraction in these clusters.

Until recently only 3 variable stars were known in NGC 6752  (Hogg
1973; Clement 1996 and references therein). One of these is a
population II Cepheid and there is insufficient information on the
other two to define their types. The horizontal branch of NGC 6752 is
blue and the cluster contains no RR~Lyr stars.  Three photometric
studies of variable and binary stars based on  HST data have been
published during the last three years.  Shara et al. (1995) reported
null results in a search for short period variables in the core region
of NGC~6752, while Bailyn et al. (1996) identified 3 candidate
cataclysmic variables, also in the core of the cluster.  From an
analysis of the broadened, asymmetric main sequence in NGC 6752,
Rubenstein \& Bailyn (1997) determined that the binary fraction is
probably in the range 15\%--38\% within the core radius.

In this contribution we present an analysis of CCD photometry of NGC
6752. This data set is best suited for a search for variable stars in
the outer parts of the cluster.  A separate paper will be devoted to
the analysis of photometry for the central part of the cluster based on
data obtained in 1998 with the 2.5-m du Pont telescope (Kaluzny et al.,
in preparation)

\section{Observations and Data Reduction} 

Time-series photometry of NGC~6752  was obtained during the interval
1996 June 23 -- 1997 September 15 with the 1.0m Swope telescope at Las
Campanas Observatory.  In 1996 a Loral CCD was used as the detector,
with a scale of 0.435 arcsec/pixel and a field of view of $14.8\times
14.8$ arcmin. Two cameras were used in 1997. The first (LCO camera
SITe1) has a field of view $23.8\times 23.8$ arcmin with a scale of
0.70 arcsec/pixel. The second  (LCO camera SITe3) has a field of view
of $14.8\times 22.8$ arcmin with a scale of 0.435 arcsec/pixel. In all
cases the observations were approximately centered on the cluster.  A
total of 539 $V$-band images, 42 $B$-band images, and 2 $U$-band images
were collected on 14 nights.  Exposure times were sufficiency long --
ranging from 100 to 300 s for the $V$-band, depending on the seeing --
to ensure accurate photometry for stars located 2-3 mag below the
cluster turnoff.  The cluster was monitored for 28h35m, 13h30m and
12h00m with the SITe3, Loral and SITe1 cameras, respectively.

Instrumental photometry was extracted using DoPHOT (Schechter, Mateo 
\& Saha 1993). We used DoPHOT in the fixed-position mode, with the stellar
positions measured on "template" images (either the best images
obtained during a given run or a combination of the 2-3 best images).
For each one of the CCD cameras used in this survey a separate data
base was constructed using procedures described in detail in Kaluzny et
al. (1996).  The total number of stars included in the $V$~filter data
bases for the SITe1, SITe3 and Loral  cameras was  45049, 43882 and
22957, respectively.  The quality of the derived photometry is
illustrated in Fig. 1 in which we have plotted the $rms$ deviation
versus average magnitude for stars measured with the SITe3 camera.
This plot includes 31442 stars with $12.1<V<20.8$ with at least 74
observations.  Photometry for stars with $V<13.5$ is poor since these
stars were frequently over-exposed.  To select potential variables we
employed three methods, as described in some detail in Kaluzny et al.
(1996).  $V$-band light curves showing possible periodic signals or smooth
changes on time scales of weeks were selected for further examination.
Eleven certain variables were identified in this way. Note that the three
variables listed in Clement (1996) are all saturated on our CCD frames.


The instrumental photometry was transformed to the standard $BV$
system using observations of standard stars from Landolt (1992),
leading to relations of the form:
\begin{eqnarray}
v = a_{1} + V + a_{2}\times (B-V) \\
b = a_{3} + B + a_{4}\times (B-V)
\end{eqnarray}

The linear coefficients $a_{2}$ and $a_{4}$ were separately derived for
each of the CCD cameras. The additive constants $a_{1}$ and $a_{3}$ were
derived based on $BV$ photometry
of 12 secondary standards selected from Cannon \& Stobie (1973). 
Average values of 
the $B-V$ color were used
when transforming the observations of the variables from instrumental 
$v$ magnitudes to standard $V$ magnitudes 
This procedure leads to systematic errors not exceeding
0.003 mag.\footnote{The values of $a_{2}$  in Eq. 1 had values
from --0.02 to 0.03 depending on the CCD. None of the detected variables
has an  observed variation of $B-V$  exceed 0.1 mag. Hence,
systematic errors due to the adoption of an average $B-V$ color  in Eq. 1
do not exceed 0.003 mag}

On the night of  Jun 26, 1997 we used the SITe1 camera to secure two
$600$-sec $U$-band frames centered on the cluster.  These frames,
supplemented with a pair of short $BV$ exposures ($V$ 25-sec, $B$
35-sec) and a pair of long exposures $BV$ exposures ($V$ 120-sec, $B$
150-sec), were used to derive $UBV$ photometry for a large sample of
stars from the cluster field. The color-magnitude diagram based on
these data is discussed in Sec. 4. The transformation from the
instrumental to the standard $UBV$ system was determined using
observations of several Landolt (1992) fields obtained over the whole
observing sub-run:
\begin{eqnarray}
v = c_{1} + V + 0.04\times (B-V) \\
b-v = c_{2}  + 0.912\times (B-V)\\
u-b = c_{3}  + 0.957\times (U-B)
\end{eqnarray}
The zero points of our $UBV$ photometry of NGC 6752 were derived 
from observations of secondary standards from Cannon \& Stobie (1973).

\section{Results for variables}

In Table 1 we list equatorial coordinates of the 11 newly identified
variables. Approximate angular distances 
of the variables from the cluster center are given in the 4th column.
The last column gives a variability type for each object
(ECL -- eclipsing binary; EW -- contact binary; EA -- detached eclipsing
binary, SX -- SX Phe variable).
The limiting radius of the cluster is estimated to be 31 $arcmin$
(Webbink 1985)
and all variables from Table 1 are located within this radius.

Variables V4, V5 and V11, which are located at large radii from the
cluster center, were outside the field of view of the SITe3 and Loral
cameras and were observed only with the SITe1 camera. Variable V10 was
outside the field of view of the Loral camera.  Variables V12 and V13,
which are located relatively close to the projected cluster center, are
absent in the data base for the SITe1 camera because of crowding. All
other variables are present in all data sets.  Finding charts for the
11 variables are given in Figs. 2 and 3.

Variable V9 is the brighter component of an unresolved blend of two
images.  However, this close visual pair could be resolved on images
obtained with the du Pont telescope. The fainter component has a
$V$-magnitude of $V=17.65 $ and color $B-V=0.45$, and we used these
values to decompose the observations of the combined  image obtained
with the Swope telescope.

Figure 4 shows the location of all of the newly identified variables on
the cluster color-magnitude diagram (CMD). For the SX~Phe stars the
plotted positions correspond to the intensity-averaged magnitudes. For
the eclipsing variables the magnitudes at maximum light are plotted.
The phased light curves of variables V4-9 and V11-14 are shown in Fig.
5. The periods of variability were derived using an algorithm based on
an {\it analysis of variance} statistic introduced by
Schwarzenberg-Czerny (1989, 1991).  Variable V10 showed flat light
curves on all but one night.  On the night of June 7 1997 that variable
showed an eclipse event lasting more than 5 hours.  Figure 6 shows
time-domain light curves of V10 obtained on the nights of June 2 1997
and June 7 1997.

\newpage

\setcounter{figure}{1}
\begin{figure}[h]
\vbox to9cm{\rule{0pt}{9cm}}
\includegraphics{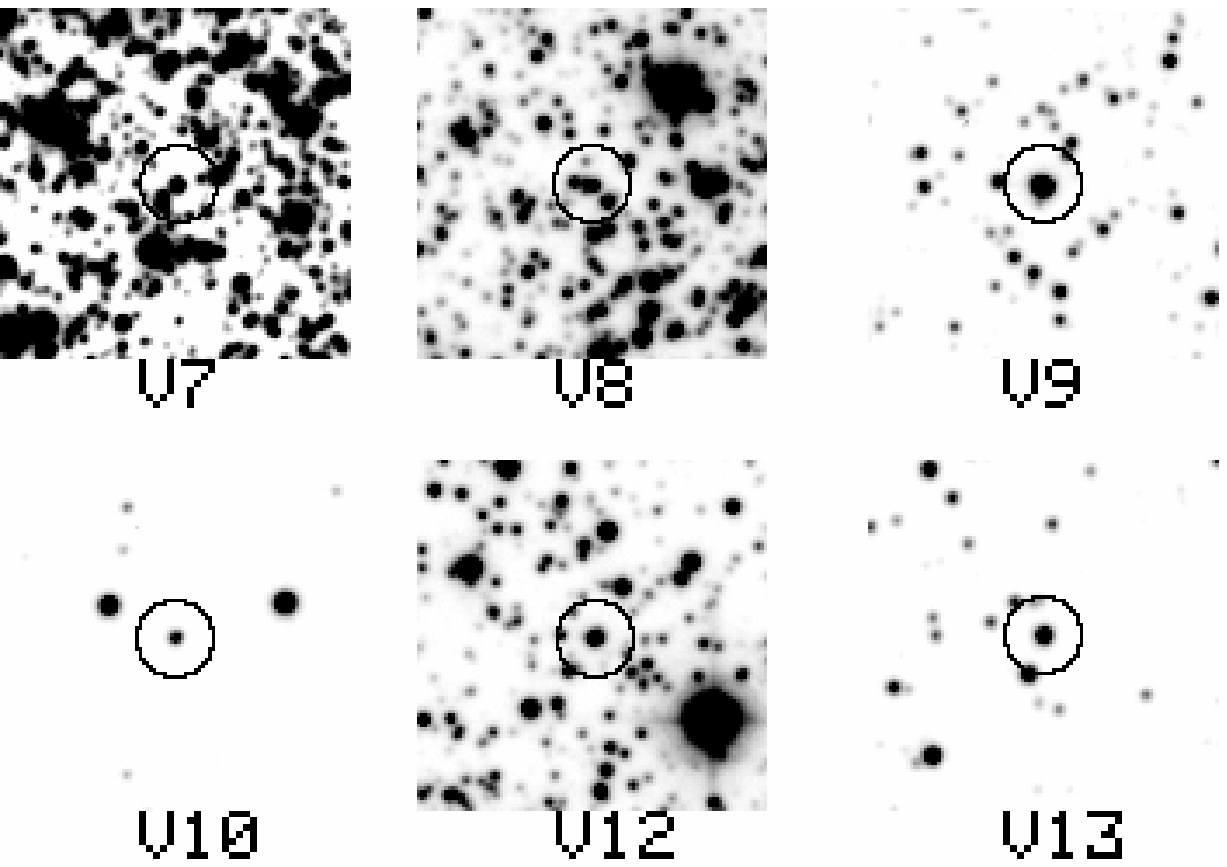}
\caption{Finder charts for variables V7-10 and V12-13. Each chart is 44
arcsec on a side with north up and east to the left.}
\end{figure}

\begin{figure}[h]
\vbox to4cm{\rule{0pt}{4cm}}
\includegraphics{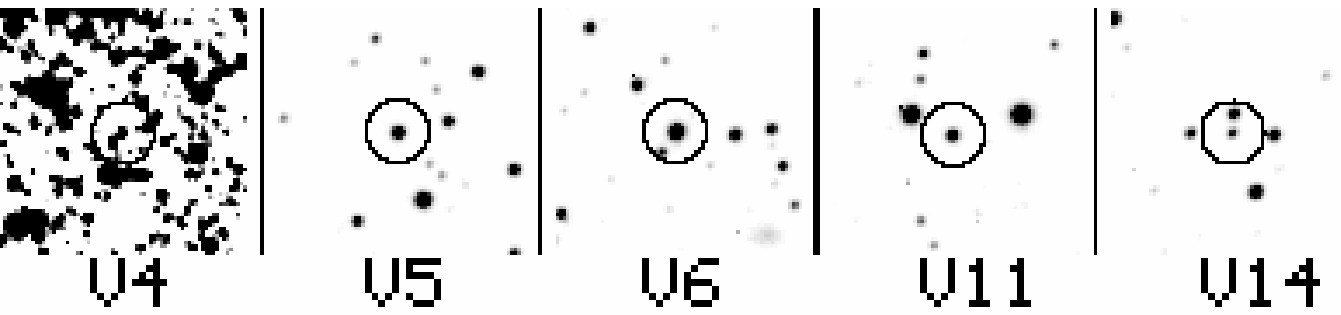}
\caption{Finder charts for variables V4-6, V11 and V14. Each chart is 44
arcsec  on a side with north up and east to the left.}
\end{figure}

\newpage

\begin{figure}[h]
\vbox to14cm{\rule{0pt}{14cm}}
\includegraphics{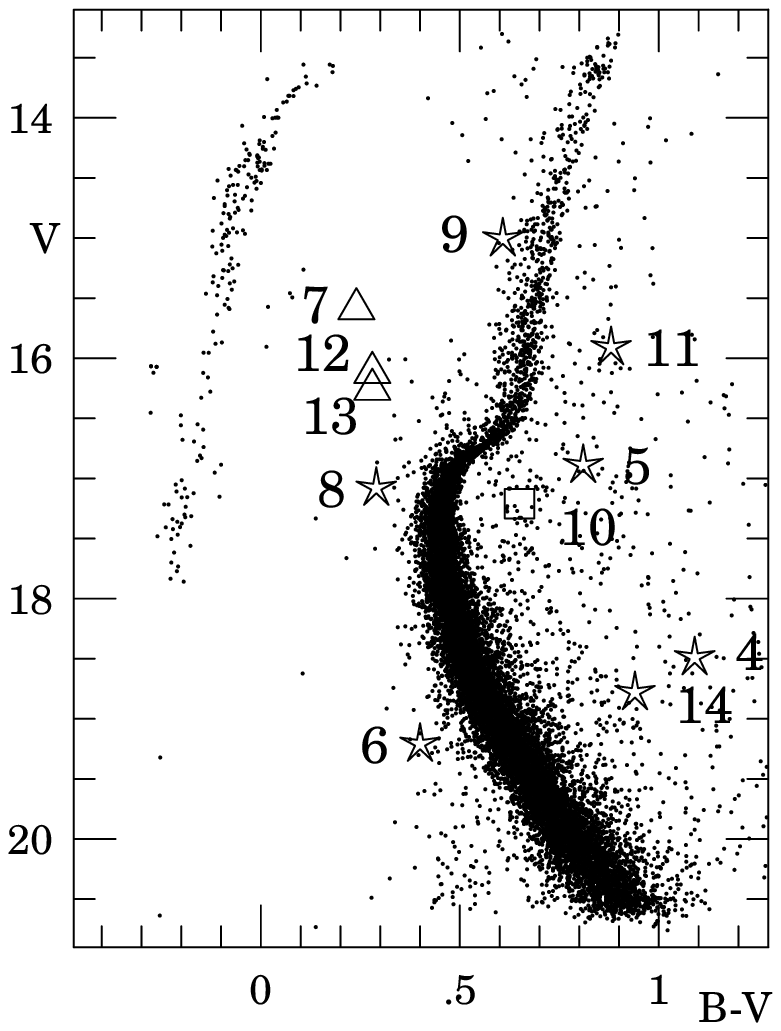}
\caption{A $V - (B-V)$ CMD for NGC~6752 with the positions of the
variables marked: triangles -- SX~Phe stars;  asterisks -- contact binaries;
open square -- probable detached binary.}
\end{figure}

\subsection{SX Phe stars}

The periods, average colors, intensity averaged $V$ brightnesses and
full  amplitudes of the 3 identified SX~Phe stars are listed in Table 2.
All three variables are  candidate  blue  straggler stars.  The observed
luminosities of these SX~Phe stars are consistent with membership in
NGC~6752. This is demonstrated in Fig. 7 which shows the positions of
the variables in a period $vs.$ absolute magnitude diagram. The standard
relation for SX Phe stars from  McNamara (1997) is also shown.
Absolute magnitudes for the SX~Phe stars were calculated assuming a
distance modulus of $(m-M)_{V}=13.02$ (Harris 1996).

While V17 is clearly a fundamental mode pulsator, based on its amplitude 
and asymmetric light curve, a classification of V12 and V13 is more
problematical. Observational error and the low amplitudes of these stars combine to complicate a determination of the asymmetry of the light curves.
In addition, as McNamara (1997) points out, classification of a star
as a first overtone pulsator based only on low amplitude and
symmetrical light curves is itself questionable. We note here simply
that the properties of these three SX Phe stars are consistent with
pulsation in the fundamental mode and membership in NGC 6752.  

\begin{figure}[h]
\vbox to15cm{\rule{0pt}{15cm}}
\includegraphics{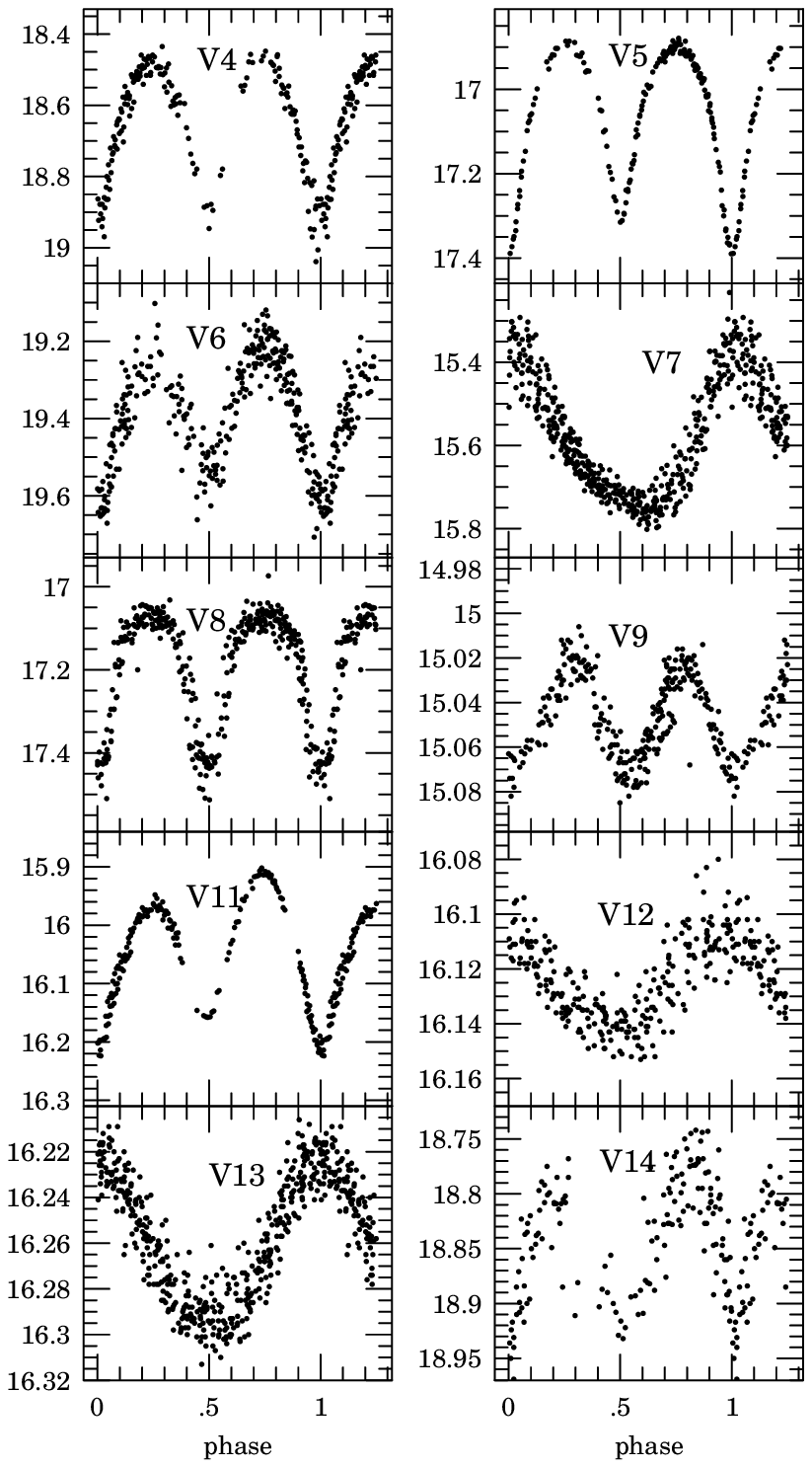}
\caption{Phased $V$-band light curves for the 10 newly identified variables in
the field of NGC~6752.}
\end{figure}

\newpage

\begin{figure}[h]
\vbox to4cm{\rule{0pt}{4cm}}
\includegraphics{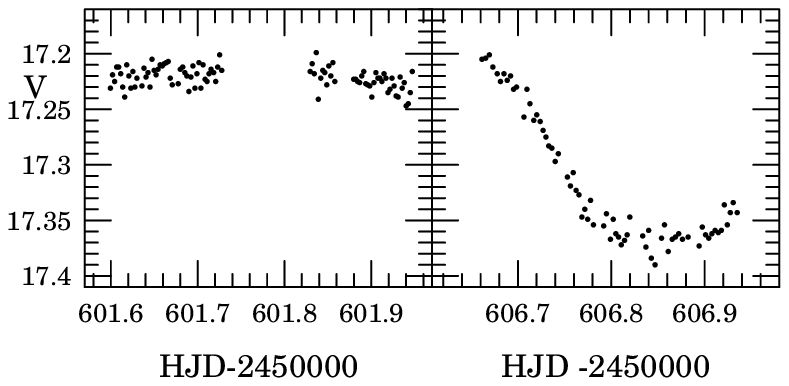}
\caption{The $V$ band light curves of variable V10 obtained on the
nights of 1997 June 2 and 7.}

\end{figure}

\begin{figure}[h]
\vbox to5cm{\rule{0pt}{5cm}}
\includegraphics{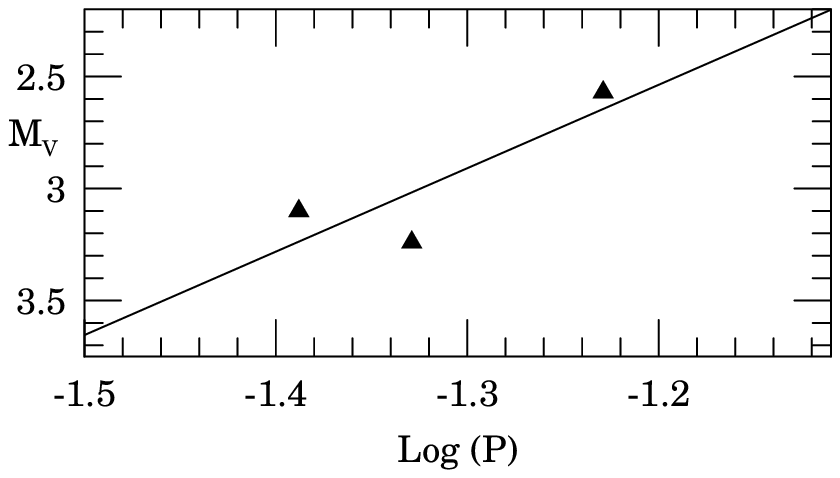}
\caption{Period vs. absolute magnitude diagram for SX~Phe stars from the field
of NGC~6752. The solid line represents the standard relation for
fundamental mode pulsators (McNamara 1997).}
\end{figure}

\subsection{Eclipsing binaries}

Our sample of newly identified variables includes 8 eclipsing
binaries. Table 3 lists some  basic photometric characteristics of
their light curves. Seven objects can be classified as contact binaries
(EW type eclipsing binaries according to the GCVS scheme) and one --
variable V10 -- is most likely a detached eclipsing binary. We have
managed to catch just one eclipse-like event for V10. Our data indicate
that the orbital period of that binary is longer than 2 days.  The
position of V10 on the cluster CMD (see Fig. 4) does not support its
membership in NGC~6752 unless the error in $B-V$ is unusually
large. In addition, the variable is located
relatively far from the cluster center. Determination of the radial
velocity and/or proper motion of V10 is necessary in order to clarify
the membership status of this potentially important binary.

Of the seven identified contact binaries only V8 is located among  the
blue stragglers on the cluster CMD. Variable V6 occupies a position
slightly below the cluster main-sequence. V11 is located about 0.2 mag
to the red of the subgiant branch. Its position on the CMD indicates
that it does not belong to the cluster.  The remaining 3 EW systems are
located to the red of the cluster main-sequence. We have applied the
absolute brightness calibration established by Rucinski (1995) to
estimate $M_{V}$ for the newly identified contact
binaries\footnote{Rucinski \& Duerbeck (1997) have derived a new
version of  the calibration $M_{V}=M_{V}(log P, B-V)$ using a sample of
EW systems with $HIPPARCOS$ parallaxes.  However, that new calibration
is based on stars from the solar neighbourhood and does not include a
metallicity term.}.  That calibration gives $M_{V}$ as a function of
period, unreddened color and metallicity:
\begin{eqnarray}
M_{V} = -2.38 log P +4.26(B-V)_{0} +0.28 -0.3[{\rm Fe/H}]
\end{eqnarray}
\noindent
where $P$ is the period in days.

We adopt $[{\rm Fe/H}]= -1.61$ and $E(B-V)=0.04$  for NGC~6752
(Harris 1996). The formal errors of the estimated values of $M_{V}$ are
about 0.5 magnitude.  Figure 8 shows a period versus apparent
distance modulus diagram for the 7 contact binaries from the cluster field.
The apparent distance modulus was calculated  as the difference between
$V_{max}$  and $M_{V}^{cal}$ for each system. The data presented in
Fig.  8 indicate that only V14 can be considered a possible cluster
member. The remaining 6 EW systems are most likely
background/foreground objects.

\begin{figure}[h]
\vbox to5cm{\rule{0pt}{5cm}}
\includegraphics{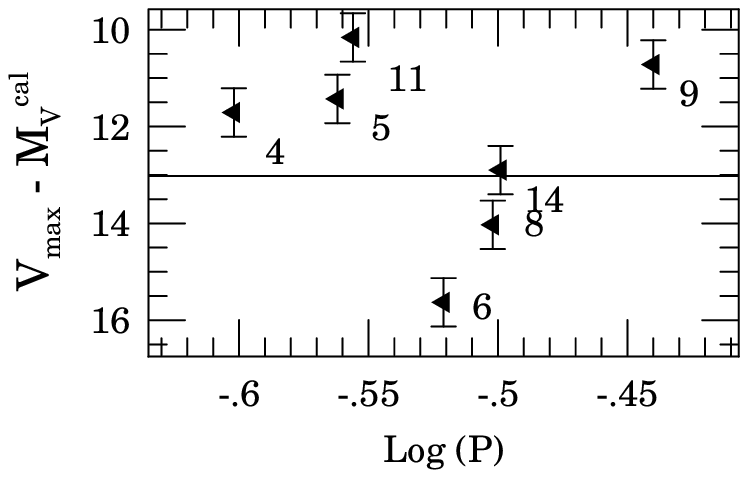}
\caption{Period vs. apparent distance modulus diagram for contact
binaries from the field of NGC~6752. The horizontal line
corresponds to a cluster distance modulus of $(m-M)_{V}=13.02$. Vertical
bars  represent formal errors of $M_{V}$
derived from the Rucinski (1995) calibration. }
\end{figure}

Although V9 is classified as a contact binary with $P=0.36$~d we cannot
exclude the possibility that the true period is  $P=0.72$~d and that
the variable is a single spotted star, presumably of type RS CVn.  Its
position on the cluster CMD suggests that in this case the star is a
subgiant belonging to the cluster (see Fig. 4).

\section{The color-magnitude diagrams}

Since the pioneering studies by Alcaino (1972) and Cannon \& Stobie
(1973) it has been known that the horizontal branch of NGC~6752 is
strongly dominated by stars located to the blue side of the instability
strip.  The CMD of the cluster was studied in detail by Buonanno et al.
(1986) based on deep photographic photometry. They showed that the
horizontal branch of NGC~6752 spans about 4 magnitudes in $V$ reaching
$V\approx 18.0$ ($M_{V}\approx 5.0$) on its faint end.  Detailed
studies of the properties of the hot subdwarfs forming extended
horizontal branch (EHB) of NGC 6752  have been published by  Heber et
al. (1986) and Moehler, Heber \& Rupprecht (1997).

\begin{figure}[h]
\vbox to11cm{\rule{0pt}{11cm}}
\includegraphics{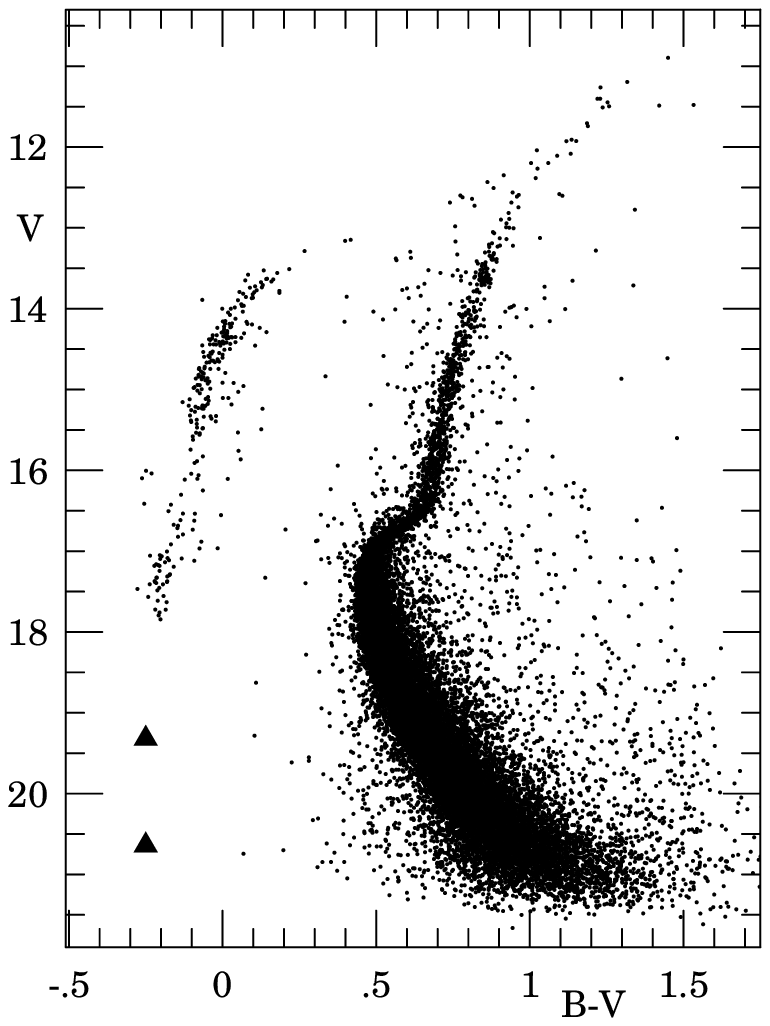}
\caption{A $V - (B-V)$ CMD for NGC~6752 based on the
template images obtained with the  SITe3 CCD camera.
Two candidate  faint EHB stars are marked with triangles.
Note also the excess of stars on the red side of the EHB. }
\end{figure}

As a by-product of our variability program we have obtained medium-deep
CMD's for the surveyed fields. In Fig. 9 we present a $V/B-V$ CMD based
on the pair of "template" images obtained with the SITe3 camera.  The
photometry was extracted using the Daophot/Allstar package (Stetson
1987).  The CMD shows several features of the EHB discussed  in some
detail by Buonanno et al. (1986).  In particular, we confirm the
presence of stars in the EHB gap between $V\approx 16.0$ and $V\approx
17.0$, and an apparent faint limit to the EHB stars at $V\approx
18.0$.  We comment briefly on two features of this CMD that bear on the
origin and evolution of EHB stars.  The first is that there are several
stars located slightly to the red of the EHB in Fig. 9. Although we
cannot exclude the possibility that some of them are field objects, we
find  that these stars are strongly concentrated toward the cluster
center. These stars are  candidate composite systems, consisting of an
EHB star plus a red dwarf.  We are planning a more detailed discussion
of these systems in a forthcoming paper based on data obtained with the
2.5m du Pont telescope (Kaluzny et al. in preparation).  The  CMD in
Fig. 9 also shows two blue stars with $B-V\approx -0.25$ which form an
apparent faint extension of the EHB of the  cluster. These stars are
marked with triangles in Fig. 9 and their coordinates as well as $UBV$
photometry are listed in Table 4.  In Fig. 10 we present a $V/U-B$ CMD
based on the data collected with the SITe1 camera. The positions of the
two faint blue stars in Fig. 9  are marked.  Faint blue stars located
below the EHB have been observed in other stellar clusters. Kaluzny \&
Rucinski (1995) noted the presence of several $UV$-bright stars with
$M_{V}>8$ in the center of NGC~6791 and more recently Cool et al.
(1998) identified similar objects in the core of NGC~6397. They argue
that these stars are either low-mass helium white dwarfs or
very-low-mass core-He-burning stars. We note that both faint  blue
objects discovered in the field of NGC~6752 are good  targets for
spectroscopic follow up with large  ground-based telescopes.

\begin{figure}[h]
\vbox to11cm{\rule{0pt}{11cm}}
\includegraphics{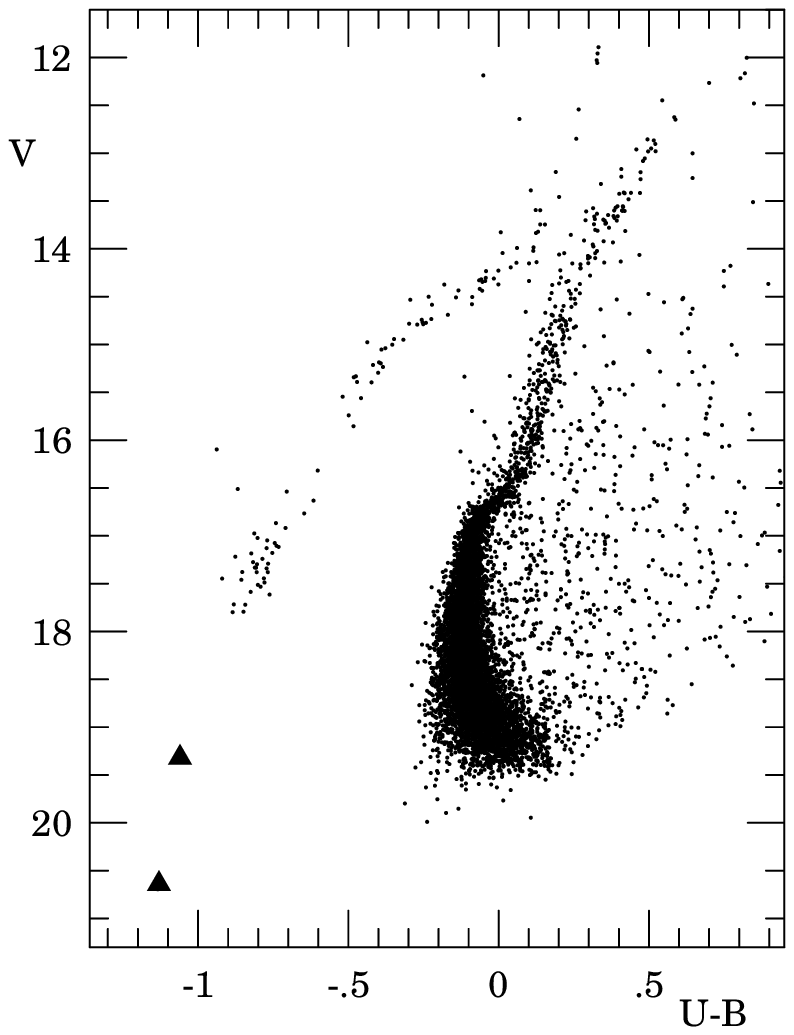}
\caption{A $V - (U-B)$ CMD for NGC~6752 based on the
images obtained with the  SITe1 CCD camera.
The two candidate faint EHB stars are marked with triangles.}
\end{figure}

Photometry and equatorial coordinates of all stars plotted in Figs. 9
and 10 are available on request from the second author of this paper.

\section{Conclusions}

We have used time series CCD observations to identify eleven new variables
in the direction of the globular cluster NGC~6752. Three of these
variables are SX Phe stars which are likely to be cluster members.
Six out of the seven identified  contact binaries are probably field
objects. One candidate detached eclipsing binary has been discovered and
follow-up observations are planned 
to get complete light curves for this potentially important variable. 
As a side-result we obtained $UBV$ photometry for a large sample
of stars from the cluster field,  and we note the presence of two faint blue
objects located below the apparent cut-off of the EHB of the cluster.

\acknowledgements
JK, WP and WK were supported by the Polish Committee of Scientific
Research through grant 2P03D-011-12 and by NSF grant AST-9528096 to
Bohdan Paczy\'nski. 
We are indebted to Dr. B. Paczy\'nski for
his long-term support to this project. Thanks are due to Randy Phelps 
for taking some 
data which were used in this paper. Dr. Dona Dinescu
kindly provided us with positional data for stars from 
the NGC~6752 field.

\vspace{7pt}


\begin{deluxetable}{lllrl}
\tablecaption{Equatorial coordinates for NGC~6752 variables. The fourth
column gives the angular distance from the cluster center.}
\tablewidth{0pt}
\tablehead{
\colhead{ID} & \colhead{RA(2000)} & \colhead{Dec(2000)} & \colhead{R} &
\colhead{Type} \nl
\colhead{} & \colhead{h:m:sec} & \colhead{deg:$\arcmin$:$\arcsec$} & 
\colhead{arcmin} & \colhead{}
}
\startdata
V4 & 19:09:23.34 & $-$60:07:02.5  & 12.0 & ECL/EW \nl
V5 & 19:09:35.77 & $-$59:49:20.4  & 12.0 & ECL/EW \nl
V6 & 19:09:52.70 & $-$59:58:11.1  & 6.6  & ECL/EW \nl
V7 & 19:10:54.21 & $-$60:00:11.9  & 1.0  & SX \nl
V8 & 19:11:10.76 & $-$59:58:52.9  & 2.0  & ECL/EW \nl
V9 & 19:11:26.37 & $-$60:01:24.2  & 4.3  & ECL/EW? \nl
V10& 19:11:28.77 & $-$59:48:25.6  & 10.1 & ECL/EA? \nl
V11& 19:12:16.44 & $-$59:53:07.8  & 10.6 & ECL/EW \nl
V12& 19:10:34.36 & $-$59:56:56.6  &  2.7 & SX \nl
V13& 19:10:14.44 & $-$60:05:03.4  & 6.7  & SX \nl
V14& 19:10:12.75 & $-$60:01:20.9  & 4.8  & ECL/EW \nl

\enddata
\end{deluxetable}



\begin{deluxetable}{clccl}
\tablecaption{Light curve parameters for SX Phe stars from NGC~6752.
$A_{V}$ is the full range of variability.}
\tablewidth{0pt}
\tablehead{
\colhead{ID} & \colhead{$P[{\rm day}]$} & 
\colhead{$<B-V>$} & \colhead{$<V>$} & \colhead{$A_V$} 
}
\startdata
V7 & 0.059076 & 0.24 & 15.59 & 0.46 \nl
V12& 0.040895 & 0.28 & 16.12 & 0.035 \nl
V13& 0.046877 & 0.28 & 16.26 & 0.07 \nl

\enddata
\end{deluxetable}



\begin{deluxetable}{clccll}
\tablecaption{Light curve parameters for eclipsing binaries from the field
of NGC~6752}
\tablehead{
\colhead{ID} & \colhead{Period} & \colhead{$T_{0} HJD+$} & 
\colhead{$V$} & \colhead{$V$} & \colhead{$B-V$} \nl
\colhead{} & \colhead{day} & \colhead{245 0000} & 
\colhead{Max} & \colhead{Min} & \colhead{Max}
}
\startdata 
V4 &0.25023   & 601.385 & 18.49 & 18.94 & 1.09 \nl
V5 &0.27362   & 601.419 & 16.89 & 17.39 & 0.81 \nl
V6 &0.30067   & 601.389 & 19.21 & 19.58 & 0.40: \nl
V8 &0.31496   & 256.609 & 17.08 & 17.42 & 0.29 \nl
V9 &0.36302   & 601.356 & 15.02 & 15.07 & 0.61 \nl
V10 &  ?      &  ?      & 17.21 & 17.37 & 0.65 \nl
V11 & 0.27833 & 601.349 & 15.91 & 16.21 & 0.88 \nl
V14 & 0.31746 & 605.580 & 18.78 & 18.91 & 0.94 \nl

\enddata
\end{deluxetable}



\begin{deluxetable}{lllrrr}
\tablecaption{Photometry and equatorial coordinates 
for faint blue stars identified in the field of NGC~6752}
\tablehead{
\colhead{ID} & \colhead{RA(2000)} & \colhead{Dec(2000)} & 
\colhead{V} & \colhead{B-V} & \colhead{U-B} \nl
\colhead{} & \colhead{hr:min:sec} & \colhead{deg:min:sec} &
\colhead{} & \colhead{} & \colhead{}
}
\startdata
B1  & 19:10:10.07 & --60:09:22.3 &  19.32 & --0.25 & --1.06       \nl
    &             &             &$\pm$0.04 &$\pm$ 0.05 & $\pm$ 0.05 \nl
B2 & 19:10:23.23 & --59:54:55.9  &  20.64 & --0.25 & --1.13       \nl
   &             &              &$\pm$ 0.11 & $\pm$ 0.15 & $\pm$ 0.15 \nl

\enddata
\end{deluxetable}


\clearpage


\newpage

\begin{thebibliography}{}
\bibitem[1972]{alcaino}
Alcaino, G. 1972, A\&A 16, 220
\bibitem[1993]{bailyn}
Bailyn, C.D., Rubenstein, E.P., Slavin, S.D.,  Cohn, H.,
Lugger, P., \& Cool, A.M. 1996, ApJL 473, L31
\bibitem[1986]{buo86}
Buonanno, R., Caloi, V., Castellani, V., Corsi, C.E., Fusi Pecci, F.,
Gratton, R. 1986, A\&AS 66, 79
\bibitem[1973]{cannon}
Cannon, R.D., Stobie, R.S. 1973, MNRAS 162, 227
\bibitem[Clement 1996]{clem}
Clement, C.M. 1996, An Update to Helen Sawyer-Hogg's Third Catalogue of
Variable Stars in Globular Clusters, private communication 
\bibitem[1998]{cool}
Cool, A.M., Grindlay, J.E., Cohn, H.N., Lugger, P.M., Bailyn, C. 1998,
ApJ 508, 75 
\bibitem[Harris 1996]{harris} 
Harris, W.E. 1996, AJ 112, 1487
\bibitem[1986]{heber86}
Heber, U., Kudritzki, R. P., Caloi, V.,
 Castellani, V., Danziger, J. 1986 A\&A 166, 369
\bibitem[Hogg 1973]{hogg73} 
Hogg, H.S. 1973, Publ. DDO 6, Nr. 3, p. 1
\bibitem[1995]{kal95}
Kaluzny, J., \& Rucinski, S.M. 1995, A\&AS 114, 1
\bibitem[1996]{kal96} 
Kaluzny, J., Kubiak, M., Szymanski, M., Udalski, A., 
Krzeminski, W., Mateo, M. 1996, A\&AS 120, 139
\bibitem[Kaluzny, Thompson \& Krzeminski 1997]{kal97}
Kaluzny, J., Thompson, I., Krzeminski, W. 1997, AJ 113, 2219
\bibitem[1992]{land}
Landolt, A. 1992, AJ 104, 340
\bibitem[1996]{macnamara}
McNamara, D.H. 1997, PASP 109, 1221
\bibitem[1997]{moe97}
Moehler, S., Heber, U., Ruprecht, G. 1997, A\&A 319, 109
\bibitem[Paczy\'nski 1997]{pacz97}
Paczy\'nski, B. 1997, in {\it The Extragalactic Distance Scale}, eds.
Livio, M., Donahue, M., \& Panagia, N., Cambridge Univ. Press,
Cambridge, p. 273
\bibitem[1997]{rub}
Rubenstein, E.P., \& Bailyn, C.D. 1997, ApJ 474, 701
\bibitem[1997]{rucinski95}
Rucinski, S.M. 1995, PASP 107, 648
\bibitem[1997]{rucinski97}
Rucinski, S.M., \& Duerbeck, H.W. 1997, PASP 109, 1340
\bibitem[1993]{schechter}
Schechter, P.L., Mateo, M., Saha, A. 1993, PASP 105, 1342
\bibitem[1993]{shara}
Shara, M.M., Drissen, L., Bergeron, L.E., \& Paresce, F. 1995, ApJ 441, 617
\bibitem[1989]{sch1}
Schwarzenberg-Czerny, A., 1989, MNRAS, 241, 153
\bibitem[1991]{sch2}
Schwarzenberg-Czerny, A., 1991, MNRAS, 253, 198
\bibitem[1987]{stet} 
Stetson, P.B. 1987, PASP 99, 191 
\bibitem[1985]{web}
Webbink, R. 1985, in {\it Dynamics of Star Clusters}, IAU Symp. 113, 
eds. J. Goodman \& P. Hut, p. 541

\end{thebibliography}
\end{document}